\documentstyle[aps,epsfig]{revtex}

\begin{document}
\title{Acoustically Induced Radiation of a Charged Particle Channeling
in a Crystal}
\draft
\author{Andrei V. Korol\cite{Email1} }
\address{Department of Physics,
St.Petersburg State Maritime Technical University,
Russia, \\198262 St. Petersburg, Russia }
\author{Andrey V. Solov'yov\cite{Email2} }
\address{
A.F.Ioffe Physical-Technical Institute of the Academy of Sciences of
Russia, \\ 194021 St. Petersburg, Russia}
\author{Walter Greiner\cite{Email3} }
\address{Institut f\"{u}r Theoretische Physik der Johann Wolfgang
Goethe-Universit\"{a}t, 60054 Frankfurt am Main, Germany}

\date{December 4, 1997}
\maketitle

\begin{abstract}
We suggest a new type of radiation (acoustically
induced radiation\, --\, AIR), which is generated by
ultrarelativistic particles channeling in crystal along a crystal
plane (or axis), which is bent by a transverse acoustic wave.
The AIR mechanism allows to make an undulator
with characteristics inaccessible in the undulators based on the motion of
particles in the periodic magnetic fields and also in the field of the
laser radiation.
The intensity of AIR can easily be made larger than the
intensity of the radiation in a linear crystal and
can be varied in a wide range by varying the frequency and the amplitude
of the acoustic wave.
\end{abstract}

\begin{figure}
\hspace{1.8cm}\epsfig{file=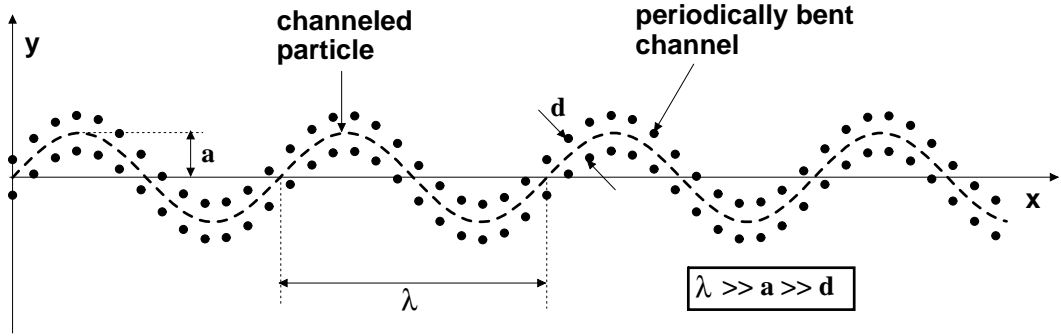,angle=270,width=14cm}
\vspace{0.5cm}
\caption{ Beam of particles channeled through the crystal bent
by the transverse accoustic wave.
}
\label{fig1}
\end{figure}

We suggest the new type of the undulator radiation (acoustically induced
radiation\, --\, AIR), which is generated by the ultrarelativistic
charged particle channeled in a crystal along a crystal axis or a
crystal plane, which is bent by a transverse acoustic wave (AW).
In this letter we consider channeling of a
positron along the plane bent by the 
transverse monochromatic standing AW
of  the frequency withinin $1-10^2$ MHz range.
Due to bending of the crystal planes caused by the
AW, the beam of  channeled positrons penetrating through the
channel oscillates in the transverse direction as illustrated
in figure \ref{fig1}. Transverse
oscillations of positrons caused by the AW become
the effective source of radiation of undulator type due to the constructive
interference of the photons emitted from the similar parts of the
trajectory. As we demonstrate below, the number of
oscillations can vary in a wide range from few up to few hundreds
oscillations per $cm$ depending on the chosen parameters.

In addition to AIR, channeled positrons emit ordinary
channeling radiation \cite{1} due to their transverse
motion inside the channel. However, the frequency of
these oscillations is much higher than the frequency of the  
transverse oscillations caused by the AW (see the estimates below). 
Therefore, two motions are well  separated, 
and the AIR mechanism can be treated independently from
the ordinary channeling radiation.
A similar situation occurs in a one-arc bent crystal,
where channeled charged particles generate
additional synchrotron type radiation due to  the curvature
of the channel \cite{2,8}.
This component of the total radiation intensity
leads to  the undulator effect in the channel periodically bent by AW.
We demonstrate that the
intensity of AIR can be made larger than the ordinary channeling radiation.
The important feature of the AIR consists in the possibility
to vary significantly the intensity and the shape of the spectral
distribution of AIR by varying the frequency and the amplitude of AW.  The
suggested mechanism of AIR allows to make an undulator with the parameters
$N$ and $p$ varying in a wide range (where $N$ is the number of periods of
the undulator and $p$ is the undulator constant) which is inaccessible in
the undulators based on the motion of charged particles in the periodic
magnetic fields and also in the field of the laser radiation \cite{3}.  In
the suggested scheme, AIR is generated by the relativistic charged
particles, with relativistic factors $\gamma=\varepsilon/mc^2$ in the range
$\gamma=1-10^7$ and above ($c$ is the velocity of light, $m$ is the mass of
the particle).  The large range of $\gamma$ available in the modern
colliders at present or in nearest future for various charged particles,
both light and heavy, together with the wide range of frequencies and the
amplitudes possible for AW in crystals allows to generate the AIR photons
with the energies up to the TeV region.

The phenomenon, which we consider in the present paper, is
an example of the charged particle channeling in a bent crystal.
The process of channeling of the charged particles in a bent crystal
is presently of  great current interest (see e.g. \cite{4,5}),
because of its possible practical applications for the manipulation
 of charged particle beams of high energy.
In these papers, both theoretical and experimental, bent crystals were
assumed or used as specially prepared or grown.
Here we suggest completely different way of bending the crystal
based  on the use of the acoustic waves.
It is important to note that the condition for channeling in a
bent crystal (see e.g. \cite{2,5}) can be fulfilled in this case too.
As a result, motion of particle in the field of the channel
bent by the transverse AW becomes periodic and thus acquires
essential features
of the motion of the particle in the periodic magnetic field.  However,
parameters of particle trajectory, such as the
amplitude or the oscillation period, are not accessible in magnetic fields
available at present in laboratory conditions.

Now let us analyze the conditions which must be fulfilled when
considering the bent crystal as an undulator.
Consider an ultra-relativistic particle
(of mass $m$, charge $q>0$, energy $\varepsilon$, relativistic factor
$\gamma \gg 1$, $v\approx c$)
channeling near a crystallographic plane.
Extension to the case of axial channeling of electrons is rather
straightforward.  
Let the transverse monochromatic AW be transmitted along
the direction of projectile's motion, and let 
$v_t$, $\nu$, $\lambda=v_t/\nu$, $a$ be the AW
phase velocity, the frequency, the wave length and the wave amplitude,
respectively.
Under the action of AW the planar channel, linear initially,
will be bent.

Bending of a channel by the AW becomes significant if
the AW amplitude becomes larger than the interplanar distance, $d$,
$d\ll a$.
In this letter we analyze the case when the shape of the channel does
not vary much during the time of flight, $\tau =L/c$, of the particle
through the crystal of the thickness $L$, i.e. $\tau$ is much smaller
than the AW period $T$:  $\tau \ll T$.
The channeling process in a bent crystal takes place if
the maximal centrifugal force in the channel, $m \gamma v^2/R_{min}$
(where $R_{min}$  is  a minimum curvature radius of the bent channel)
is less the force caused by 
the maximal interplanar field $U_{max}^{\prime}$
\cite{2,5}:
\begin{equation}
m \gamma v^2/R_{min}  <  q\,  U_{max}^{\prime}.
\label{1}
\end{equation}

Provided (1) is fulfilled, the projectile, which enters the
crystal under the angle $\theta$ much less than the critical angle
$\theta_{L}$, will move, being trapped in the channel, along the
trajectory, which represents the instant shape of the accoustically
bent channel (see figure 1):

\begin{equation}
y(x)=a\sin\left(2\pi{x \over \lambda }\right),
\qquad x=[0\dots L]
\label{2}
\end{equation}

The frequency of the transverse oscillatory motion of a positron
moving along the trajectory (\ref{2}), $2\pi c/\lambda$, is much smaller
than the frequency of the oscillatory motion  inside the
channel, $2\pi\sqrt{2U_{max}/m\gamma} /d$ \cite{1}, provided condition
(\ref{1}) is fulfilled and if $a\gg d$. 
The frequencies
of the two types of motions are significantly different so that they
can be treated independently.

The minimum curvature radius of the trajectory (\ref{2}) is equal to
$R_{min}=(\lambda/2\pi)^2/a$.
Thus, decreasing $\lambda$ and increasing $a$,
we decrease $R_{min}$ and  increase the maximum acceleration of
the particle in the channel.
As a result, photon emission due to the projectile's acceleration in
 the bent channel may  be significantly enhanced.
Below we demonstrate that this radiation, which is emitted coherently from
similar parts of the trajectory, may dominate considerably over the
radiation caused by the acceleration of the particle in the linear
channel.

For the trajectory (\ref{2}) the relation (\ref{1}) reads as
\begin{equation}
\nu^2\, a < C \equiv \gamma^{-1}\cdot
\left({ v_t \over 2\pi} \right)^2 \cdot
\left({ q\, U_{max}^{\prime} \over mc^2}\right)
\label{4}
\end{equation}
\noindent
and determines ranges of $\nu$, $a$ and $\gamma$ for which the
channeling process as well as the undulator radiation, can occur for
given crystal and crystallographic plane (the parameters
$U_{max}^{\prime}$ and $v_t$ are subject
to the choice of a particular crystal and a plane)
and for given projectile type, characterized by a rest
mass and  a charge.

Both the motion of the projectile in the bent channel and
the spectrum of the generated electromagnetic radiation are of the
undulator-type, only if $\lambda \ll L$, i.e. if  the
channeling particle oscillates many times within the length $L$ of
the crystal.
The suggested undulator, 'AW + channeling particle', is
characterized as any other undulator \cite{3} by the frequency,
$\omega_0= 2\pi\, c/\lambda$,
 and the undulator parameter, $p = 2\pi\gamma\, a/\lambda$.

\begin{figure}
\hspace{1.8cm}\epsfig{file=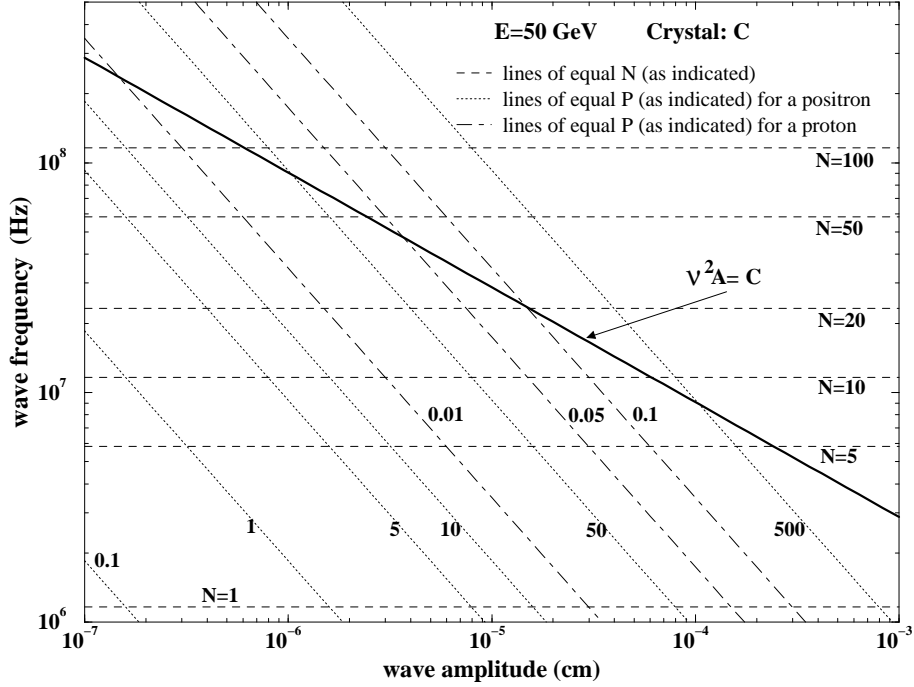,angle=270,width=12cm}
\vspace{0.5cm}
\caption{
The ranges of $\nu$ (in $Hz$) and  $a$ (in $cm$)
in which the channeling process is possible for 50 GeV
positron and proton in a carbon crystal.
See also explanations in the text.
}
\label{fig1a}
\end{figure}

\begin{figure}
\hspace{1.8cm}\epsfig{file=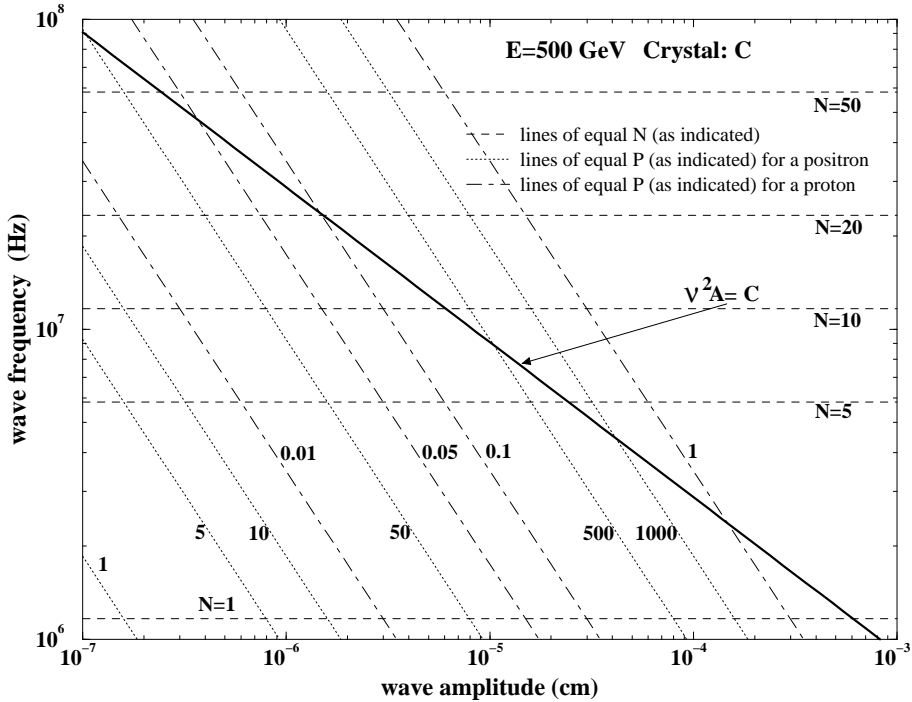,angle=270,width=12cm}
\vspace{0.5cm}
\caption{Same as for  Fig.\ 2 
but for 500 GeV projectiles}
\label{fig1b}
\end{figure}

Figs.\ \ref{fig1a}--\ref{fig1b} illustrate the ranges of
$\nu$  and  $a$  in which the channeling process
for a positron and a proton in a carbon crystal is possible.
The cases \ref{fig1a} and \ref{fig1b} correspond to the energies
$50$ and $500$ $GeV$.
The solid thick line in both figures represents the boundary
$\nu^2\, a = C$ (see (\ref{4})).
In each figure the range of validity of (\ref{4}) lies below this line.
Dotted and dashed-dotted lines indicate the constant values of the
undulator parameter $p$ for  a positron and a  proton.
The dashed lines correspond the constant values of the
number of the AW periods per 1 cm:  $N\, ({\rm cm^{-1}}) = \lambda^{-1}$.
All data presented in Figs.\ \ref{fig1a}--\ref{fig1b} and subsequently
refer to a projectile channeling near the $(110)$ crystallographic plane.
We use  $v_t$ ($10^5 cm/s$): 11.64,  2.81 \cite{6} and
$U_{max}^{\prime}$ ($Gev/cm$): 12.0,  43.0 for
C and W crystals, respectively \cite{5}.


Figs.\ \ref{fig1a}--\ref{fig1b} demonstrate that the parameters $p$
and  $N$ vary in  wide ranges: $N=1...100$, $p=0.1-500$ for
projectile positron and  $p= 0.001-0.1$ for a proton.
These values are by more than an order of magnitude larger
than those accessible in the undulators based on the motion of
the charged particles either in periodic magnetic fields or 
in the field of the laser radiation \cite{3}.

In the limit  $N,p \gg 1$ one can calculate the spectral intensity of
radiation (per 1 cm) emitted by a projectile positron moving along
the path (\ref{2}), by utilizing the following formula
deduced from general expression, given in \cite{7},
which has been obtained within the framework of quasi-classical
approximation:
\begin{equation}
{d \varepsilon \over d (\hbar \omega)\, L}
=
\alpha N p\, {\omega^{\prime} \over \omega}\, \left(
G_1(y) + \left[1+{u^2 \over 2(1+u)}\right] G_2(y)\right)
\label{5}
\end{equation}
where $\alpha$ is the fine structure constant, $\omega$ is the photon
frequency, $\omega^{\prime}=\omega\, (1+u)$, and the parameter
$u=\hbar\omega /(\varepsilon - \hbar\omega)$ takes into account the
correction due to the radiative recoil. The parameter $y$ is defined
as
$y=\left( \omega^{\prime} /  \omega_0 \gamma^2 p\right)^{2/3}$
with $\omega_0 = 2\pi\, c/\lambda$.
The functions $G_{1,2}(y)$ are
\begin{eqnarray}
G_1(y)&=& - 2 y^{5/2} \int_1^{\infty}
dx \left[\pi - \arccos\left(1-{2 \over x^3}\right)
\right] {\rm Ai}(y x)
\nonumber \\
G_2(y)&=& - 8 y^{1/2} \int_0^{\infty}
{ d\xi \over ({\rm ch}\xi)^{5/3}}\,
{\rm Ai}^{\prime}(y ({\rm ch}\xi)^{2/3})
\label{7}
\end{eqnarray}
where ${\rm Ai}(z),{\rm Ai}^{\prime}(z)$ are the Airy function and
its derivative respectively.

\begin{figure}
\hspace{1.8cm}\epsfig{file=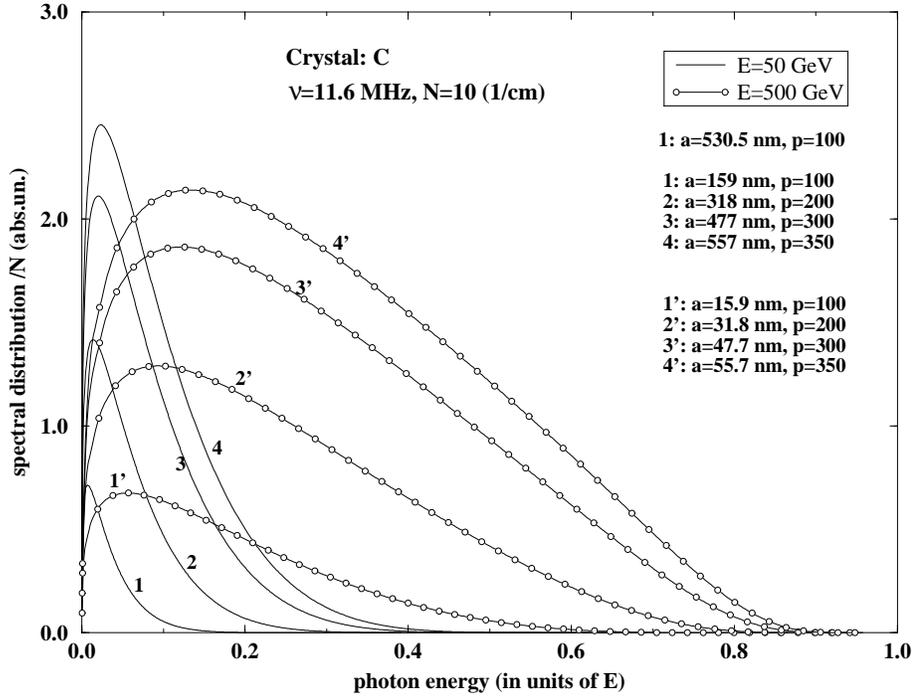,angle=270,width=12cm}
\vspace{0.5cm}
\caption{
The spectral intensity (per one period) of the AIR
emitted by a  $50$ and $500$ $GeV$  positron in a carbon
crystal calculated for the fixed AW frequency
(as indicated) and for various
parameters $a$ and $p$ (as indicated).
See also explanations in the text.}
\label{fig2a}
\end{figure}

\begin{figure}
\hspace{1.8cm}\epsfig{file=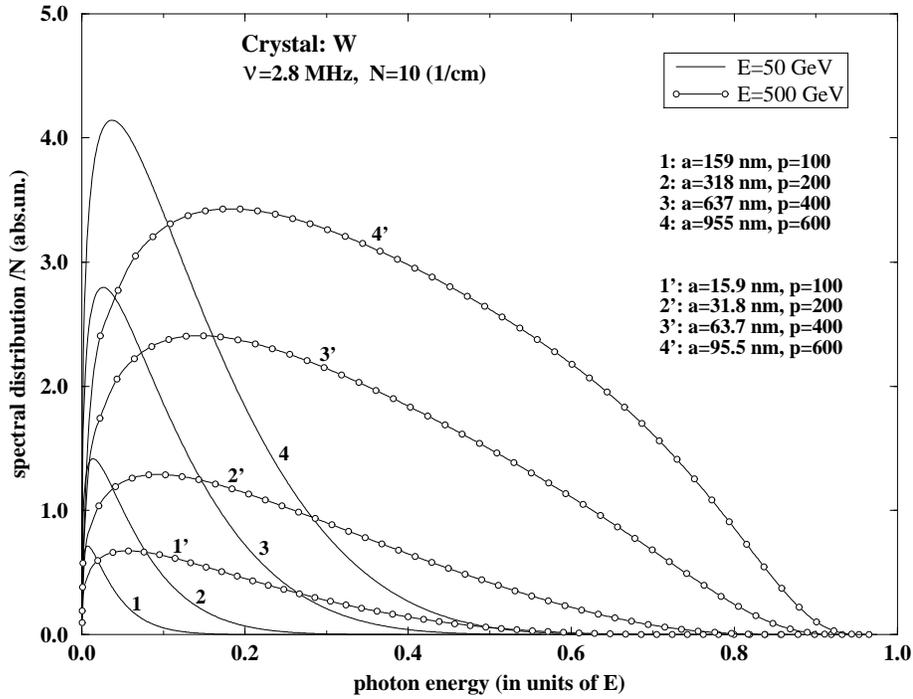,angle=270,width=12cm}
\vspace{0.5cm}
\caption{Same as for  Fig.\ 4 
but for a tungsten crystal.}
\label{fig2b}
\end{figure}

Using  (\ref{5})--(\ref{7}), we have calculated the spectral distributions of
AIR at different parameters.
The spectral distributions (per $cm$) of the radiation emitted by a
$50$ and $500$ $GeV$  positron moving along the trajectory (\ref{2})
in carbon and  tungsten crystals are plotted for various values of the
undulator parameter $p$ and the amplitude $a$ in Figs.\
\ref{fig2a}--\ref{fig2b}.
These figures demonstrate that for a fixed frequency $\nu$ the
intensity of radiation can be varied in a wide range by altering the
AW amplitude, which is proportional to the value of the undulator
parameter $p$.
These figures illustrate as well the dependence of the  spectral 
distributions on the energy of the particle.
Comparison of the spectra presented in Figs.\
\ref{fig2a}--\ref{fig2b} with those in the
case of channeling in the corresponding linear crystal \cite{1}
shows that the intensity
of AIR can be made much larger than the corresponding intensity of
radiation in the linear crystal case
by choosing the appropriate AW parameters. 

\begin{figure}
\hspace{1.8cm}\epsfig{file=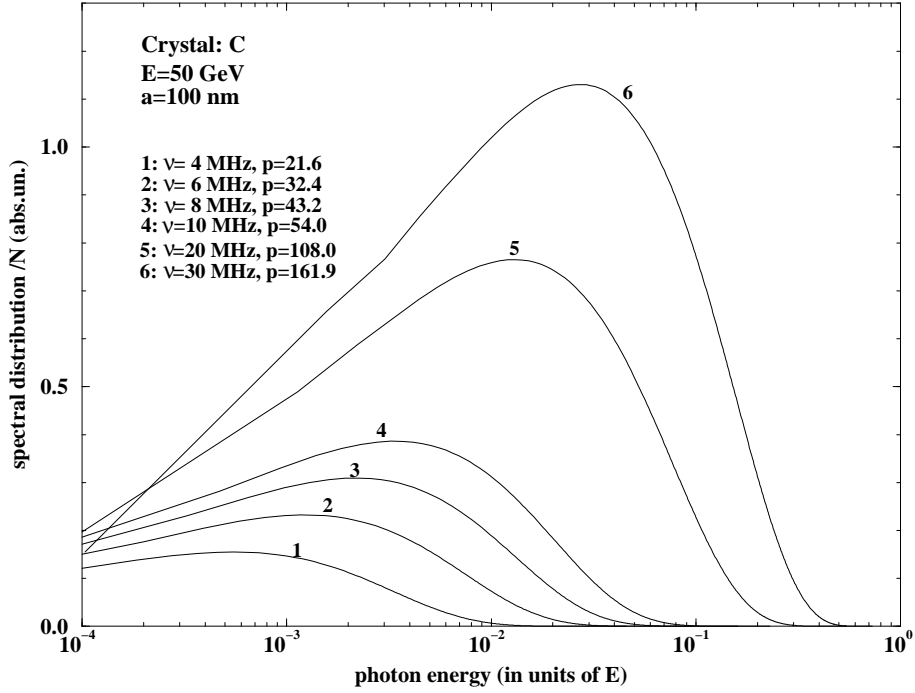,angle=270,width=12cm}
\vspace{0.5cm}
\caption{
The spectral intensity (per one period) of the AIR
emitted by a $50$ $GeV$  positron in a carbon
crystal calculated for the fixed AW amplitude (as indicated)
and for various
parameters $\nu$ and $p$ (as indicated).
See also explanations in the text.}
\label{fig3a}
\end{figure}

\begin{figure}
\hspace{1.8cm}\epsfig{file=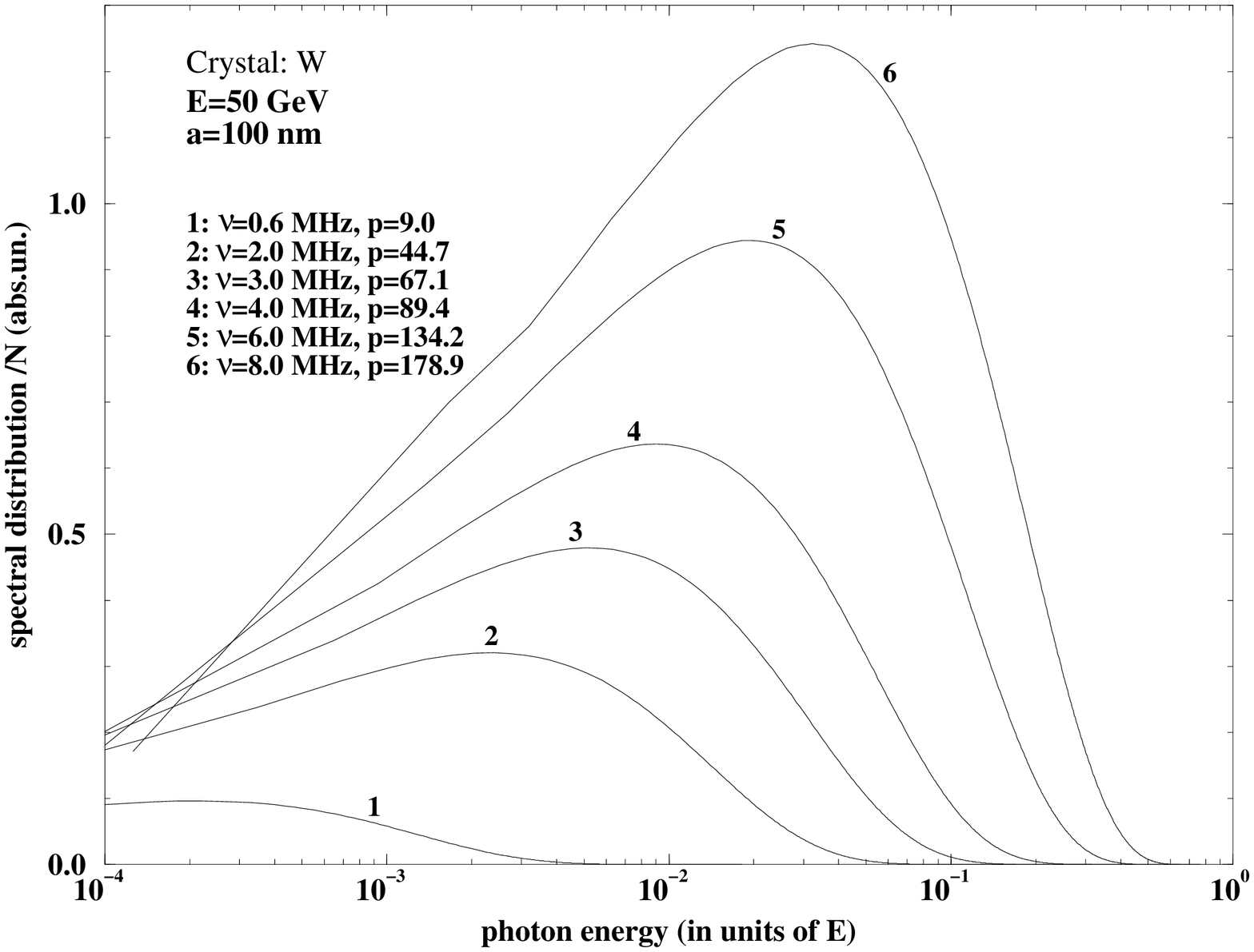,angle=270,width=12cm}
\vspace{0.5cm}
\caption{Same as for Fig.\ 6 
but for a tungsten crystal.}
\label{fig3b}
\end{figure}

The spectral intensities (per 1 $cm$) of the radiation emitted by a
$50$ $GeV$  positron moving along the trajectory (\ref{2}) are compared
for $C$ and  $W$ crystals in Figs.\ \ref{fig3a} and \ref{fig3b}.
In this calculation the AW amplitude is fixed at
100 $nm$. Other parameters are as indicated.
These figures show that the properties of the AIR radiation
depend also on the type of the material the undulator is
made from.
By varying the material one can achieve various radiation
intensities at the same parameters of the AW.

Finally, let us estimate the stability of the suggested undulator.
From general theory of undulators one can deduce \cite{1} that
the relative deviation of the  undulator resonance frequency
$\Delta \omega /\omega$ is proportional to the relative variation of
the undulator parameters, in our case to $\Delta a/a$ and $\Delta
\lambda/\lambda$.  For the resonant undulator frequencies and the
parameters considered above, the ratio
$\Delta \omega /\omega \sim 0.1$.
This means that the fluctuation of $a$ and
$\lambda$ in the AW on the level of 10$\%$ or less does not
influence much the stabilty of the suggested undulator even in the region
of very high frequencies.  Note that the relative
variation of the AW amplitude during the time of flight of the positron
through the crystal is much lower, $\Delta a/a \sim \nu L/c\sim 1/300$ at
$\nu \sim 100MHz$ and $L\sim 1cm$.

Our investigation shows that the described phenomenon can be
used for the construction of an undulator with  variable
parameters for the generation of high energy photons in
a wide range. We have discussed the plane channeling of positrons combined
with one of the most simple examples of AW.  However, other cases of AW
(longitudinal waves, spherical waves, non-monochromatic waves and various
combinations thereof), interacting with the beam of the channeled
particles (positrons, electrons, heavy ions) as well as the case of axial
channeling of electrons are worthy to study.  Another interesting question,
being raised by our work, is the possibility of the stimulated photon
emission in the undulator described above (free electron laser type). Such
work is in progress.

The authors acknowledge support from the 
DFG, GSI and BMBF.

\end{document}